\documentclass[11pt]{article}
\usepackage{moriond,epsfig}

\def\Journal#1#2#3#4{{#1} {\bf #2}, #3 (#4)}

\def\PRL{\em Phys. Rev. Lett.}

\def\D{\displaystyle}

\def\be{\begin{equation}}
\def\ee{\end{equation}}
\def\bea{\begin{eqnarray}}
\def\eea{\end{eqnarray}}

\def\to{\rightarrow}

\def\MeV{\ifmmode {\mathrm{\ Me\kern -0.1em V}}\else
                   \textrm{Me\kern -0.1em V}\fi}
\def\GeV{\ifmmode {\mathrm{\ Ge\kern -0.1em V}}\else
                   \textrm{Ge\kern -0.1em V}\fi}
\def\COM{\ensuremath{\mbox{centre--of--mass}}}
\begin{document}
\vspace*{4cm}
\title{STANDARD MODEL FITS}

\author{ A. KR\"UGER }

\address{DESY Zeuthen, Platanenalle 6,\\
15738 Zeuthen, GERMANY}

\maketitle\abstracts{
Recent results of tests of the Standard Model of electroweak interactions
are presented. Data are used from the four LEP experiments, ALEPH, DELPHI, L3,
OPAL, the SLD experiment at SLC, the TEVATRON $p \bar p$ experiments CDF and D0
and the NuTeV neutrino experiment. $\chi^{2}$-fits are performed in order to
study the consistency of the Standard Model of electroweak interactions.}

\section{Measurements}\label{sec:measurements}

Precision measurements are used for fits in the framework of the Standard Model (SM).
Measurements from LEP, SLD, TEVATRON and NuTeV are used to obtain the following
observables and pseudo-observables \cite{LEP}:

\begin{itemize}
 \item{LEP I:}
 \begin{itemize}
  \item{$Z$ lineshape: $m_{\mathrm{Z}}$, $\Gamma_{\mathrm{Z}}$,
                       $\sigma_{\mathrm{had}}^{\mathrm{0}}$, $R_{l}^{\mathrm{0}}$,
                       $A_{\mathrm{FB}}^{\mathrm{0},l}$}
  \item{$\tau$ polarisation: $P_{\mathrm{\tau}}(\cos\theta) \to {\cal A}_{\mathrm{e}},
                                 {\cal A}_{\mathrm{\tau}}$}
  \item{heavy flavour: $R_{\mathrm{b}}^{\mathrm{0}}$, $R_{\mathrm{c}}^{\mathrm{0}}$,
               $A_{\mathrm{FB}}^{\mathrm{0,b}}$, $A_{\mathrm{FB}}^{\mathrm{0,c}}$}
  \item{jet charge asymmetry: $Q_{\mathrm{FB}} \to
                      \sin^{\mathrm{2}}\theta_{\mathrm{eff}}^{\mathrm{lep}}$}
 \end{itemize}
 \item{SLD: L-R asymmetries $\to {\cal A}_{l}, {\cal A}_{\mathrm{b}}, {\cal A}_{\mathrm{c}}$}
 \item{TEVATRON/LEP II: $m_{\mathrm{t}}$, $m_{\mathrm{W}}$}
 \item{NuTeV: $\sin^{\mathrm{2}}\theta_{\mathrm{W}}\equiv 1-m_{\mathrm{W}}^{\mathrm{2}}/
              m_{\mathrm{Z}}^{\mathrm{2}}$}
\end{itemize}

The sample contains precision measurements of electroweak parameters measured at the Z peak
as well as measurements from \COM{} energies above the Z peak.\\
The final results of the $Z$ lineshape analysis and asymmetries are listed in table~\ref{tab:zline}.
The leptonic coupling parameters, ${\cal A}_{l}$ are derived from LEP I data. The combination of the results,
assuming lepton universality, yields ${\cal A}_{l}^{\mathrm{LEP}}=0.1482\pm 0.0026$ \cite{LEP}.
The result is compared to the direct measurement of ${\cal A}_{l}$ via the leptonic left-right asymmetry from SLD,
${\cal A}_{l}^{\mathrm{SLD}}=0.1513\pm 0.0021$ \cite{LEP}.

\begin{table}[htb!]
  \caption{The final $Z$ lineshape and asymmetry measurements from LEP with the quoted total uncertainties.
  \label{tab:zline}}
  \vspace{0.4cm}
 \begin{center}
  \begin{tabular}{|c|c|} 
   \hline
    $m_{\mathrm{Z}}$                     & $91187.5 \pm 2.1 \MeV$\\ 
    $\Gamma_{\mathrm{Z}}$                & $2495.2 \pm 2.3 \MeV$\\
    $\sigma_{\mathrm{had}}^{\mathrm{0}}$ & $41.540 \pm 0.037$ nb\\
    $R_{l}^{\mathrm{0}}$                 & $20.767 \pm 0.025$\\
    $A_{\mathrm{FB}}^{0,l}$              & $0.0171 \pm 0.0010$\\
   \hline
  \end{tabular}
 \end{center}
\end{table}

\begin{figure}[htb!]
  \begin{center}
    \includegraphics[width=0.5\textwidth]{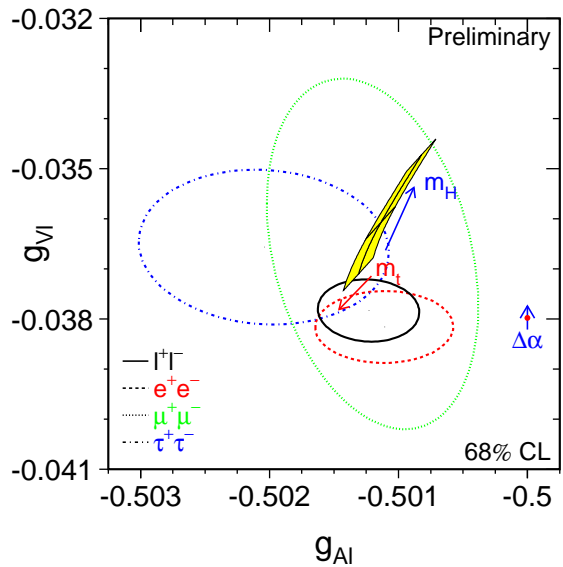}
  \end{center}
  \caption{The result of the leptonic measurements carried out at the Z peak at LEP I and SLD.
           The result is given in terms of the vector and axial-vector couplings,
           $g_{\mathrm{V}}^{l}$ and $g_{\mathrm{A}}^{l}$, of the $Z$ boson to the single lepton
           flavours and all lepton flavours combined. The result favours a low value for the mass
           of the SM Higgs boson.
  \label{fig:mhlept}}
\end{figure}

The LEP I and SLD combined result is
${\cal A}_{l}=0.1501\pm 0.0013$ \cite{LEP}. The coupling parameters are functions of the vector and
axial-vector couplings
${\cal A}_{l}=2g_{\mathrm{V}}^{l}g_{\mathrm{A}}^{l}/({(g_{\mathrm{V}}^{l})}^{2}+{(g_{\mathrm{A}}^{l})}^{2})$
The results are obtained for the separate lepton flavours and shown in figure~\ref{fig:mhlept} in terms of the vector
and axial-vector couplings, $g_{\mathrm{V}}^{l}$ and $g_{\mathrm{A}}^{l}$. The results agree well with each other
and are consistent with the hypothesis of lepton universality.\\
The arrows shown in the picture point towards higher values of the denoted
parameters (the top mass, the Higgs mass). The top mass is varied for $m_{\mathrm{t}}=174.3\pm 5.1 \GeV{}$ and the
Higgs mass is varied from $M_{\mathrm{H}}=114.1 \GeV{}$ \cite{Hoffman:2000}, the lower limit obtained from direct Higgs
searches performed at LEP II, to $M_{\mathrm{H}}=1000 \GeV{}$.\\
Particularly, a low value for the mass of the SM Higgs boson is favoured by all lepton flavours.

The heavy flavour results of LEP I, $R_{\mathrm{b}}^{\mathrm{0}}$, $R_{\mathrm{c}}^{\mathrm{0}}$,
$A_{\mathrm{FB}}^{\mathrm{0,b}}$ and $A_{\mathrm{FB}}^{\mathrm{0,c}}$, and
the final result of ${\cal A}_{\mathrm{b}}$ from SLD are listed in table~\ref{tab:had}.

\begin{table}[htb!]
 \caption{The results of $R_{\mathrm{b}}^{\mathrm{0}}$, $R_{\mathrm{c}}^{\mathrm{0}}$,
          $A_{\mathrm{FB}}^{\mathrm{0,b}}$ and $A_{\mathrm{FB}}^{\mathrm{0,c}}$ from LEP I and
          the final result of ${\cal A}_{\mathrm{b}}$ from SLD.
 \label{tab:had}}
 \vspace{0.4cm}
 \begin{center}
  \begin{tabular}{|c|c|c|} 
   \hline
    LEP I & $R_{\mathrm{b}}^{\mathrm{0}}$    & $0.21646 \pm 0.00065$\\ 
          & $R_{\mathrm{c}}^{\mathrm{0}}$    & $0.1719 \pm 0.0031$\\
          & $A_{\mathrm{FB}}^{\mathrm{0,b}}$ & $0.0994 \pm 0.0017$\\
          & $A_{\mathrm{FB}}^{\mathrm{0,c}}$ & $0.0685 \pm 0.0034$\\
   \hline
    SLD   & ${\cal A}_{\mathrm{b}}$          & $0.922\pm 0.020$\\
   \hline
  \end{tabular}
 \end{center}
\end{table}

\begin{figure}[htb!]
 \begin{center}
    \begin{tabular}{cc}
     \includegraphics[width=0.5\textwidth]{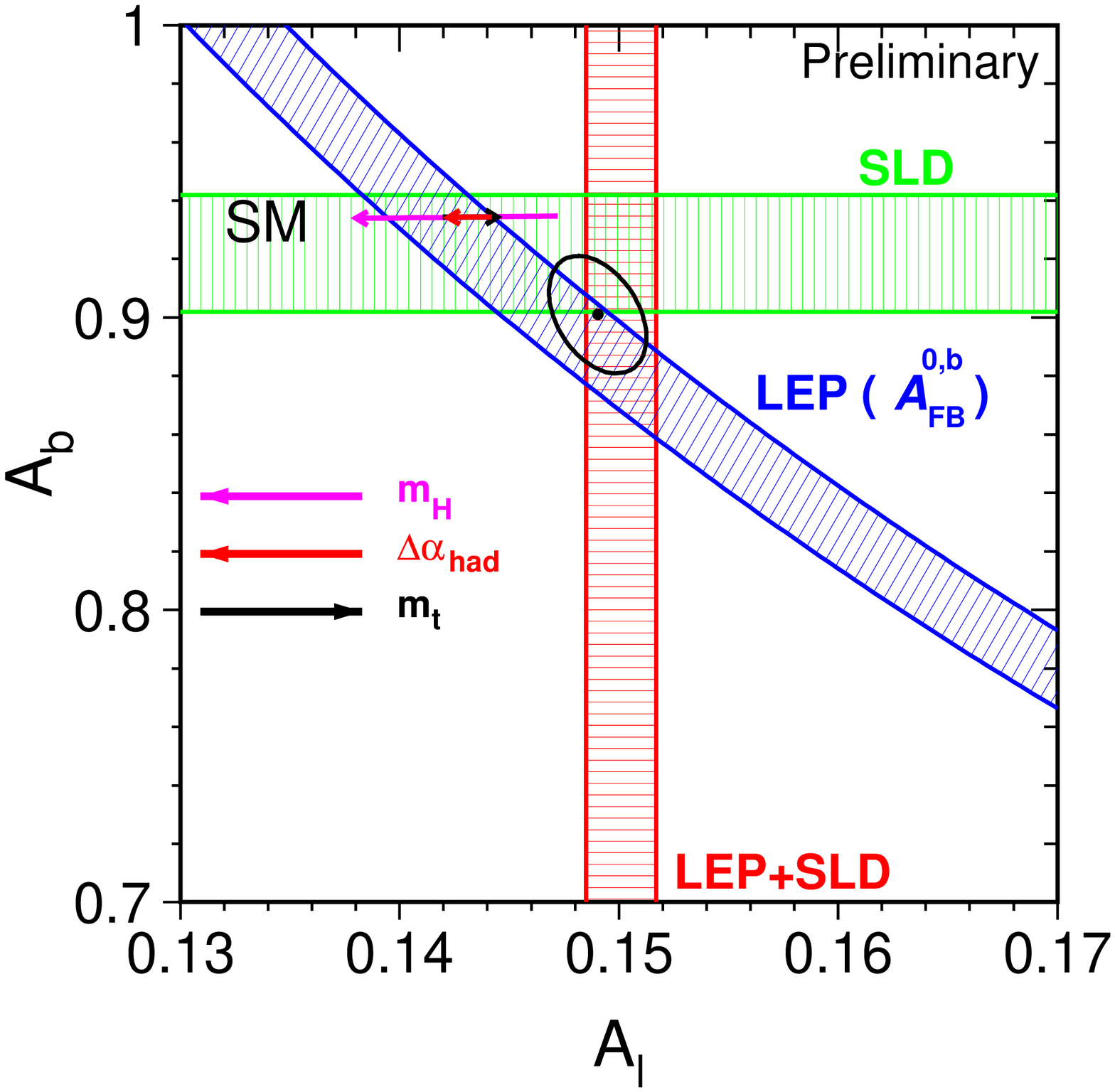} & 
     \includegraphics[width=0.5\textwidth]{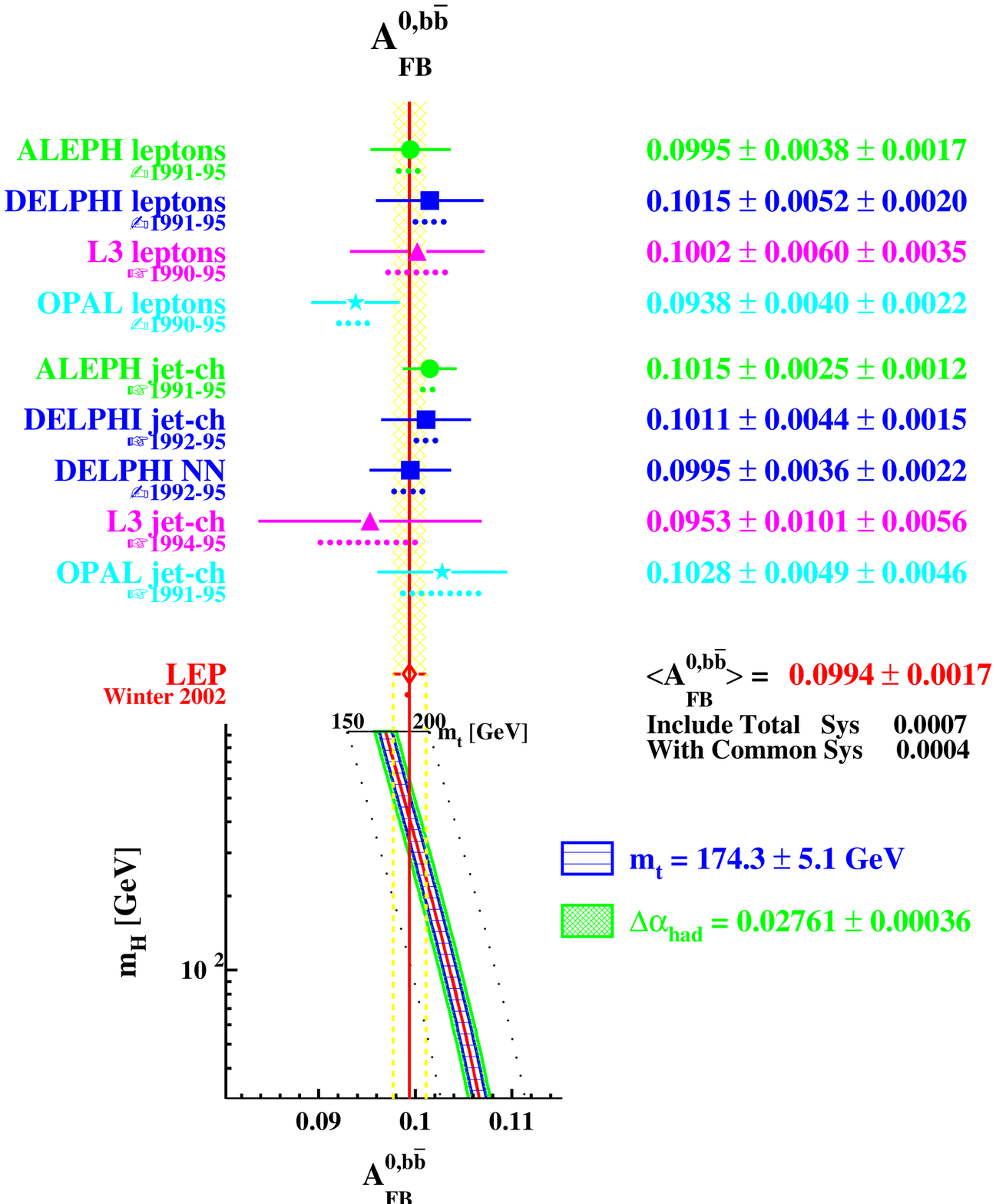} \\
    \end{tabular}
 \end{center}
  \vspace{-0.4cm}
  \caption{Left: the hadronic results depicted in the ${\cal A}_{\mathrm{b}}$-${\cal A}_{l}$ plane.
           The long arrow pointing to the left, denotes the SM dependence of the variation of the Higgs mass and the small
           arrow pointing to the right, denotes the SM dependence of the variation of the top quark mass.
           Right: the Higgs mass as a function of $A_{\mathrm{FB}}^{\mathrm{0,b}}$.
  \label{fig:mhhad}}
\end{figure}

In figure~\ref{fig:mhhad} the results of ${\cal A}_{l}$, ${\cal A}_{\mathrm{b}}$ and $A_{\mathrm{FB}}^{\mathrm{0,b}}$
is shown in the ${\cal A}_{\mathrm{b}}$-${\cal A}_{l}$ plane. The 68\% C.L. contour of the combined fit of
${\cal A}_{l}$, ${\cal A}_{\mathrm{b}}$ and $A_{\mathrm{FB}}^{\mathrm{0,b}}$ is compared to the SM prediction given
by the arrows.
The combination of $A_{\mathrm{FB}}^{\mathrm{0,b}}$ measurements from the LEP collaborations is also shown in
figure~\ref{fig:mhhad}. The combined value is compared to the SM prediction of the Higgs mass as function of
$A_{\mathrm{FB}}^{\mathrm{0,b}}$. The low value of $A_{\mathrm{FB}}^{\mathrm{0,b}}$ favours a
large Higgs mass value.

\section{Standard Model Fits}\label{sec:smfits}

\begin{figure}[htb!]
 \begin{center}
    \begin{tabular}{cc}
     \includegraphics[width=0.5\textwidth]{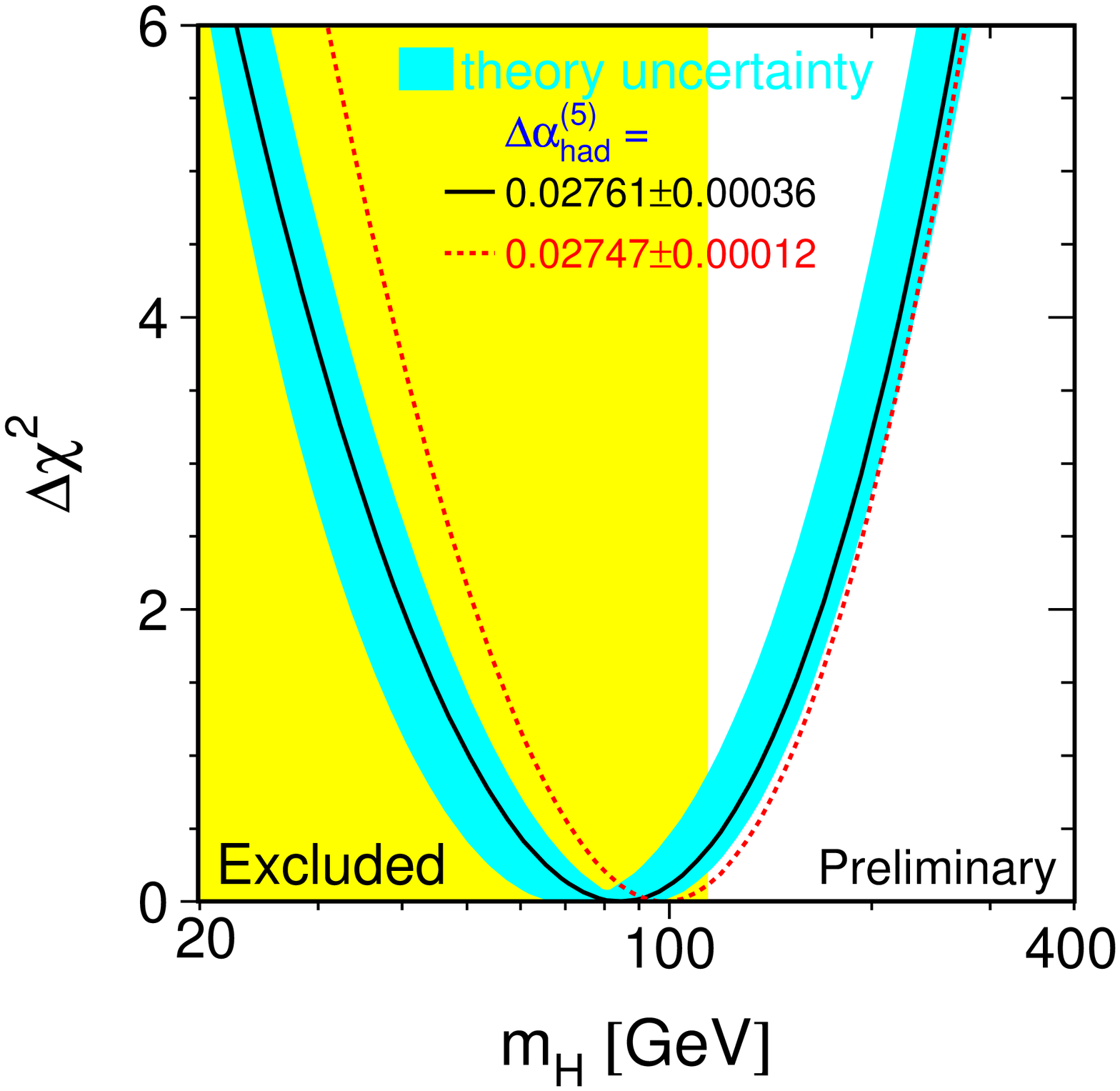} & 
     \includegraphics[width=0.4\textwidth]{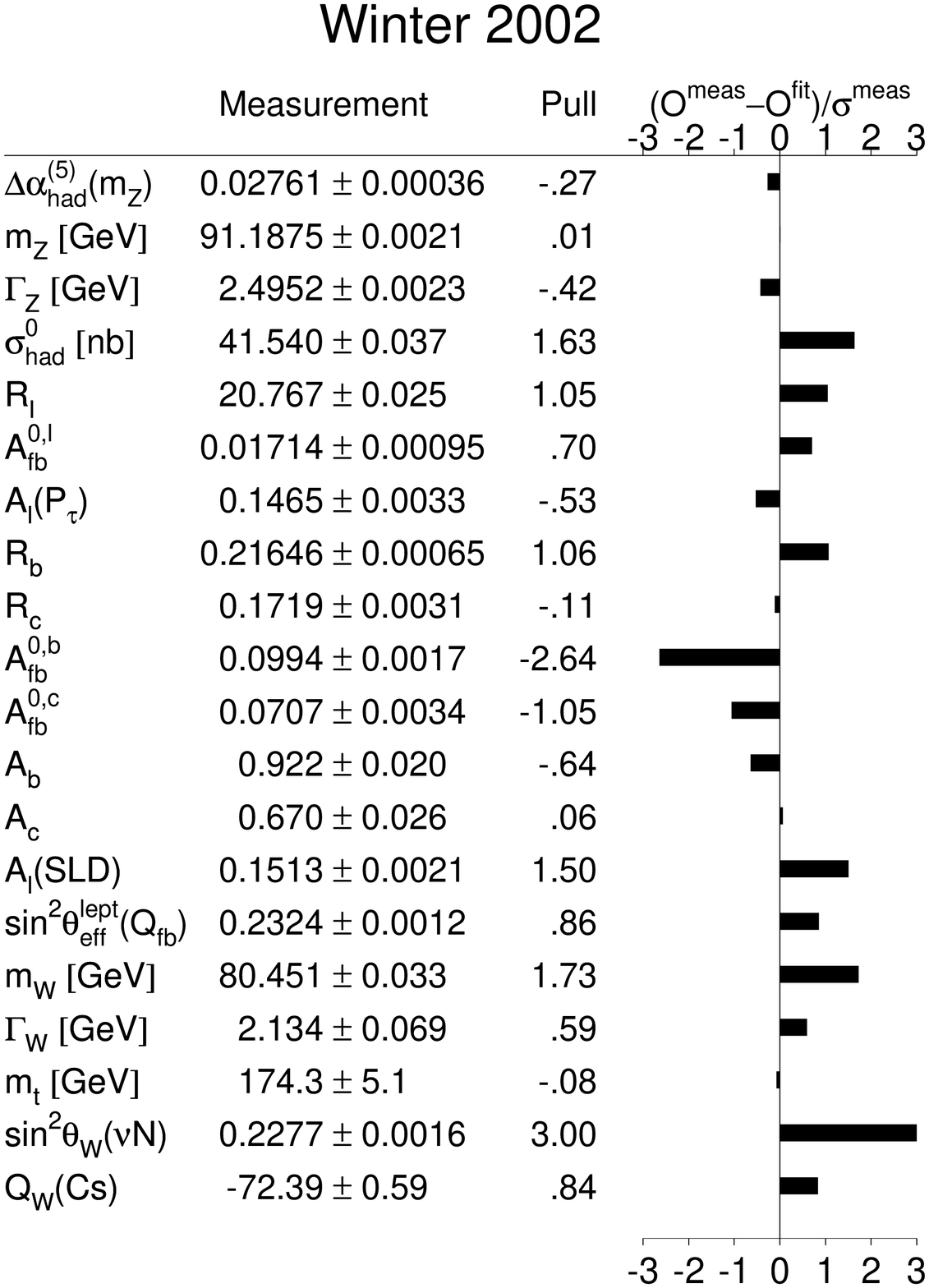} \\
    \end{tabular}
 \end{center}
  \caption{Left: $\Delta\chi^{2}$ as function of the Higgs mass.
           Right: the global fit parameters depicted in terms of pull. The definition of the pull
           is given on the head of the picture.
  \label{fig:mhglobal}}
\end{figure}

A fit to all measurements is performed for the Higgs mass. The result obtained is
$M_{\mathrm{H}}=85_{\D -34}^{\D +54} \GeV{}$.
The result is shown in figure~\ref{fig:mhglobal} for $\Delta\chi^{2}$ as a
function of the Higgs mass. The one-sided upper limit is $M_{\mathrm{H}}<196 \GeV{}$ at 95\% C.L.,
whereas the direct searches at LEP II yield a lower limit of $M_{\mathrm{H}}>114.1 \GeV{}$ \cite{Hoffman:2000}.\\
A global fit to all measurements is shown in figure~\ref{fig:mhglobal}.
The largest deviations are found for the results of $A_{\mathrm{FB}}^{\mathrm{0,b}}$ measured at
LEP I, which is 2.6 standard deviations below the SM prediction and the result
of $\sin^{\mathrm{2}}\theta_{\mathrm{W}}$ from NuTeV \cite{Zeller:2001hh}, which is 3.0 standard
deviations above the SM prediction.

\section{Conclusions}

Except for, possibly, $A_{\mathrm{FB}}^{\mathrm{0,b}}$ \cite{LEP} and the NuTeV $\sin^{\mathrm{2}}\theta_{\mathrm{W}}$
measurement \cite{Zeller:2001hh}, the SM generally describes the data well.
Leptonic data favours a low Higgs mass of ${\cal O}(20-30 \GeV)$ \cite{Chanowitz:2001bv}
compared to the direct lower limit from LEP II \cite{Hoffman:2000}.
Hadronic data (dominated by $A_{\mathrm{FB}}^{\mathrm{0,b}}$) instead tends to favour
$M_{\mathrm{H}}= {\cal O}(300-400\GeV)$.

\section*{Acknowledgements}

I wish to thank the LEP Electroweak Working Group and the LEP collaborations ALEPH, DELPHI, L3 and OPAL 
for discussions and the results, and J.H.Field for discussions.

\end{document}